\documentclass[epjST]{svjour}
\usepackage{graphicx}
\usepackage{multirow}
\usepackage{amsmath}
\usepackage{amssymb}
\usepackage{cite}
\begin{document}

\newcommand{\Emuca}{E_{\Delta}}

\title{%
  Scaling and universality in the phase diagram of the 2D Blume-Capel model
}

\author{
     J. Zierenberg\inst{1,4}
\and N. G. Fytas\inst{2,4}\fnmsep\thanks{\email{nikolaos.fytas@coventry.ac.uk}}
\and M. Weigel\inst{1,2,4}
\and W. Janke\inst{1,4}
\and A. Malakis\inst{2,3}
}
\institute{
     Institut f\"ur Theoretische Physik, Universit\"at Leipzig, Postfach 100\,920, D-04009 Leipzig, Germany
\and Applied Mathematics Research Centre, Coventry University, Coventry, CV1 5FB, United Kingdom
\and Department of Physics, Section of Solid State Physics, University of Athens, Panepistimiopolis, GR 15784 Zografou, Greece
\and Doctoral College for the Statistical Physics of Complex Systems,  Leipzig-Lorraine-Lviv-Coventry $({\mathbb L}^4)$
}

\abstract{%
  We review the pertinent features of the phase diagram of the zero-field Blume-Capel
  model, focusing on the aspects of transition order, finite-size scaling and
  universality. In particular, we employ a range of Monte Carlo simulation methods to
  study the 2D spin-1 Blume-Capel model on the square lattice to investigate the
  behavior in the vicinity of the first-order and second-order regimes of the
  ferromagnet-paramagnet phase boundary, respectively. To achieve high-precision
  results, we utilize a combination of (i) a parallel version of the multicanonical
  algorithm and (ii) a hybrid updating scheme combining Metropolis and generalized
  Wolff cluster moves. These techniques are combined to study for the first time the
  correlation length of the model, using its scaling in the regime of second-order
  transitions to illustrate universality through the observed identity of the
  limiting value of $\xi/L$ with the exactly known result for the Ising universality
  class.
}
\maketitle

\section{Introduction}
\label{secIntroduction}

The Blume-Capel (BC) model is defined by a spin-1 Ising Hamiltonian with a single-ion
uniaxial crystal field anisotropy~\cite{blume,capel}. The fact that it has been very
widely studied in statistical and condensed-matter physics is explained not only by
its relative simplicity and the fundamental theoretical interest arising from the
richness of its phase diagram, but also by a number of different physical
realizations of variants of the model, ranging from multi-component fluids to ternary
alloys and $^{3}$He--$^{4}$He mixtures~\cite{lawrie}. Quite recently, the BC model
was invoked by Selke and Oitmaa in order to understand properties of
ferrimagnets~\cite{selke-10}.

The zero-field model is described by the Hamiltonian
\begin{equation}\label{eqHamiltonian}
  \mathcal{H}
  =-J\sum_{\langle ij\rangle}\sigma_{i}\sigma_{j}+\Delta\sum_{i}\sigma_{i}^{2}
  =E_{J}+\Delta \Emuca,
\end{equation}
where the spin variables $\sigma_{i}$ take on the values $-1, 0$, or $+1$,
$\langle ij\rangle$ indicates summation over nearest neighbors only, and $J>0$ is the
ferromagnetic exchange interaction. The parameter $\Delta$ is known as the
crystal-field coupling and it controls the density of vacancies ($\sigma_{i}=0$). For
$\Delta\rightarrow -\infty$, vacancies are suppressed and the model becomes
equivalent to the Ising model. Note the decomposition on the right-hand side of
Eq.~(\ref{eqHamiltonian}) into the bond-related and crystal-field-related energy
contributions $E_J$ and $E_\Delta$, respectively, that will turn out to be useful in
the context of the multicanonical simulations discussed below.

Since its original formulation, the model \eqref{eqHamiltonian} has been studied in
mean-field theory as well as in perturbative expansions and numerical simulations for
a range of lattices, mostly in two and three dimensions, see, e.g.,
Refs.~\cite{fytas_BC,zierenberg2015}. Most work has been devoted to the
two-dimensional model, employing a wide range of methods including real space
renormalization~\cite{berker1976rg}, Monte Carlo (MC) simulations and MC
renormalization-group
calculations~\cite{landau1972,kaufman1981,selke1983,selke1984,landau1986,xavier1998,deng2005,silva2006,hurt2007,malakis,kwak2015},
$\epsilon$-expansions~\cite{stephen1973,chang1974,tuthill1975,wegner1975}, high- and
low-temperature series expansions~\cite{fox1973,camp1975,burkhardt1976} and a
phenomenological finite-size scaling (FSS) analysis~\cite{beale1986}. In the present
work, we focus on the nearest-neighbor square-lattice case and use a combination of
multicanonical and cluster-update Monte Carlo simulations to examine the first-order
and second-order regimes of the ferromagnet-paramagnet phase boundary. One focus of
this work is a study of the correlation length of the model, a quantity which to our
knowledge has not been studied before in this context. We locate transition points in
the phase diagram of the model for a wide temperature range, thus allowing for
comparisons with previous work. In the second-order regime, we show that the
correlation-length ratio $\xi/L$ for finite lattices tends to the exactly known value
of the 2D Ising universality class, thus nicely illustrating universality.

The rest of the paper is organized as follows. In Sec.~\ref{secLiterature} we briefly
review the qualitative and some simple quantitative features of the phase diagram in
two dimensions. Section~\ref{secNumerical} provides a thorough description of the
simulation methods, the relevant observables and FSS analyses. In Sec.~\ref{secFSS}
we use scaling techniques to elucidate the expected behaviors in the first-order
regime as well as the universality of the exponents and the ratio $\xi/L$ for the
parameter range with continuous transitions. In particular, here we demonstrate Ising
universality by the study of the size evolution of the universal ratio
$\xi/L$. Finally, Sec.~\ref{secSummary} contains our conclusions.

\section{Phase diagram of the Blume-Capel model}
\label{secLiterature}

\begin{figure}
  \centering
  \includegraphics{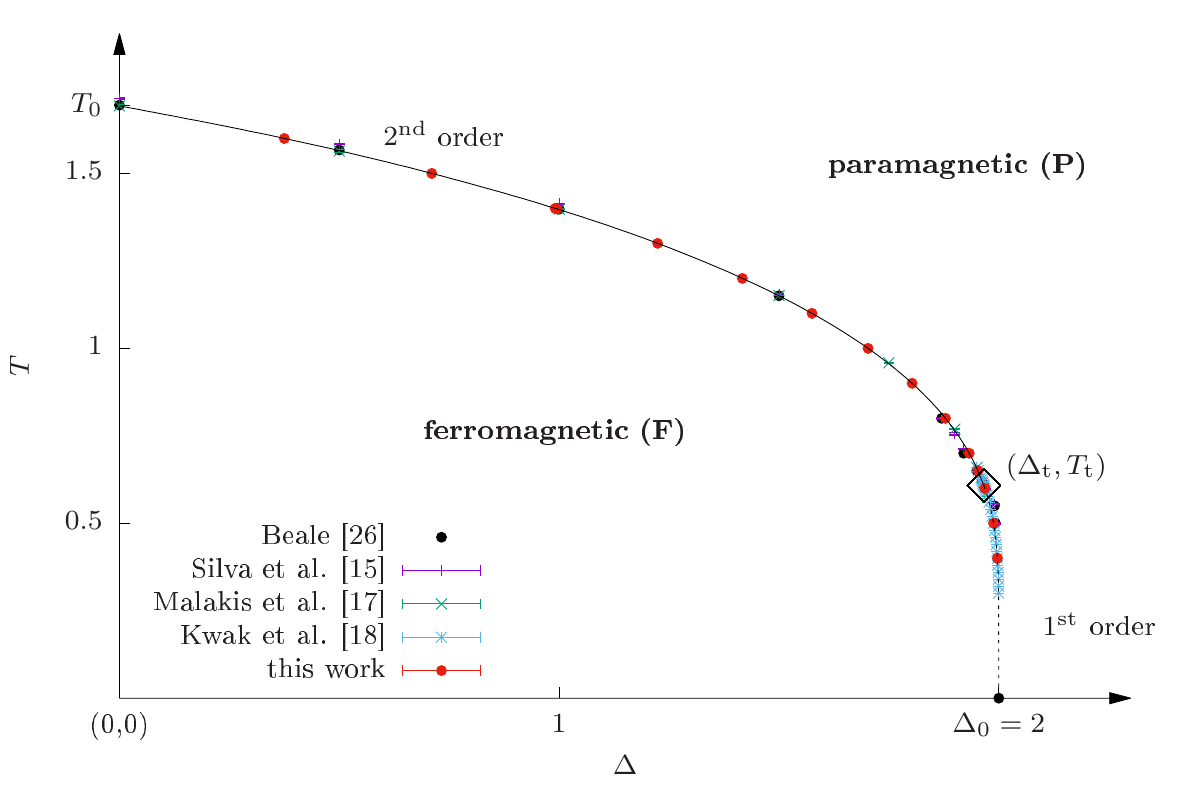}
  \caption{\label{figPhaseDiagram} %
    Phase diagram of the square-lattice, zero-field BC model in the $\Delta$--$T$
    plane.  The phase boundary separates the ferromagnetic (F) phase from the
    paramagnetic (P) phase. The solid line indicates continuous phase transitions and
    the dotted line marks first-order phase transitions. The two lines merge at the
    tricritical point $(\Delta_{\rm t}, T_{\rm t})$, as highlighted by the black
    diamond. The data shown are selected estimates from previous studies as well as
    the present work. The numerical values of all individual estimates are summarized
    in Table~\ref{tabResults} below.
  }
\end{figure}

The general shape of the phase diagram of the model is that shown in
Fig.~\ref{figPhaseDiagram}. While this presentation, comprising selected previous
results~\cite{beale1986,silva2006,malakis,kwak2015} together with estimates from the
present work, is for the square-lattice model, the general features of the phase
diagram are the same for higher dimensions also \cite{blume,capel}.  The phase
boundary separates the ferromagnetic (F) from the paramagnetic (P) phase. The
ferromagnetic phase is characterized by an ordered alignment of $\pm 1$ spins. The
paramagnetic phase, on the other hand, can be either a completely disordered
arrangement at high temperature or a $\pm1$-spin gas in a $0$-spin dominated
environment for low temperatures and high crystal fields.
At high temperatures and low crystal fields, the F--P transition is a continuous
phase transition in the Ising universality class, whereas at low temperatures and
high crystal fields the transition is of first order~\cite{blume,capel}. The model is
thus a classic and paradigmatic example of a system with a tricritical point
$(\Delta_{\rm t},T_{\rm t})$ \cite{lawrie}, where the two segments of the phase
boundary meet. At zero temperature, it is clear that ferromagnetic order must prevail
if its energy $zJ/2$ per spin (where $z$ is the coordination number) exceeds that of
the penalty $\Delta$ for having all spins in the $\pm 1$ state. Hence the point
$(\Delta_0=zJ/2,T=0)$ is on the phase boundary \cite{capel}. For zero crystal-field
$\Delta$, the transition temperature $T_0$ is not exactly known, but well studied for
a number of lattice geometries.

In the following, we consider the square lattice and fix units by choosing $J=1$ and
$k_{\rm B}=1$. The estimates shown in Fig.~\ref{figPhaseDiagram} for this case are
based on phenomenological FSS using the transfer matrix for systems up to size
$L=10$~\cite{beale1986}, standard Wang-Landau simulations up to
$L=100$~\cite{malakis}, two-parameter Wang-Landau simulations up to
$L=16$~\cite{silva2006} and $L=48$~\cite{kwak2015}, as well as the results of the
present work, using parallel multicanonical simulations at fixed temperature up to
$L=128$ ($T>T_{\rm t}$) and $L=96$ ($T<T_{\rm t}$). A subset of these results is
summarized in Table~\ref{tabResults} for comparison. Note that in the multicanonical
simulations employed in the present work we fix the temperature $T$ while varying the
crystal field $\Delta$ \cite{zierenberg2015}, whereas crossings of the phase boundary
at constant $\Delta$ were studied in most other works. In general, we find excellent
agreement between the recent large-scale simulations. Some deviations of the older
results, especially in the first-order regime, are probably due to the small system
sizes studied. We have additional information for $T=0$ where $\Delta_0 = zJ/2 = 2$
and for $\Delta=0$, where results from high- and low-temperature series expansions
for the spin-1 Ising model provide
$T_0=1.690(6)$~\cite{fox1973,camp1975,burkhardt1976}, while phenomenological
finite-size scaling yields $T_0\simeq1.695$~\cite{beale1986}, one-parametric
Wang-Landau simulations give $T_0=1.693(3)$~\cite{malakis}, and two-parametric
Wang-Landau simulations arrive at $T_0=1.714(2)$~\cite{silva2006}. Overall, the first three
results are in very good agreement. The deviations observed for the result of the
two-parametric Wang-Landau approach \cite{silva2006} can probably be attributed to
the relatively small system sizes studied there.
Determinations of the location of the tricritical point are technically demanding as
the two parameters $T$ and $\Delta$ need to be tuned simultaneously. Early attempts
include MC simulations, $[\Delta_{\rm t}=1.94,T_{\rm
t}=0.67]$~\cite{landau1972,selke1983,selke1984}, and real-space
renormalization-group calculations, $[\Delta_{\rm t}=1.97, T_{\rm
t}=0.580]$~\cite{berker1976rg,burkhardt1976rg,burkhardt1977rg,kaufmann1981rg,yeomans1981rg}.
More precise and mostly mutually consistent estimates were obtained by
phenomenological finite-size scaling,
$[\Delta_{\rm t}=1.9655(50),T_{\rm t}=0.610(5)]$~\cite{beale1986}, MC
renormalization-group calculations,
$[\Delta_{\rm t}=1.966(15),T_{\rm t}=0.609(3)]$~\cite{landau1986}, MC simulations
with field mixing,
$[\Delta_{\rm t}=1.9665(3), T_{\rm t}=0.608(1)]$~\cite{wilding1996} and
$[\Delta_{\rm t}= 1.9665(3),T_{\rm t}=0.608(1)]$~\cite{plascak2013}, transfer matrix
and conformal invariance,
$[\Delta_{\rm t}= 1.965(5),T_{\rm t}=0.609(4)]$~\cite{xavier1998}, and two-parametric
Wang-Landau simulations,
$[\Delta_{\rm t}=1.966(2),T_{\rm t}=0.609(3)]$~\cite{silva2006} and
$[\Delta_{\rm t}=1.9660(1),T_{\rm t}=0.6080(1)]$~\cite{kwak2015}.

Below the tricritical temperature, $T<T_{\rm t}$, or for crystal fields
$\Delta > \Delta_{\rm t}$, the model exhibits a {\em first-order} phase transition. This is
signaled by a double peak in the probability distribution of a field-conjugate
variable. This is commonly associated with a free-energy barrier and the
corresponding interface tension. Finite-size scaling for first-order transitions
predicts a shift of pseudo-critical points according to~\cite{binder}
\begin{equation}\label{eqShiftFirst}
  \Delta_{L}^{\ast} = \Delta^{\ast} + a L^{-D},
\end{equation}
where $\Delta^{\ast}$ denotes the transition field in the thermodynamic limit and $D$
is the dimension of the lattice. Note that a completely analogous expression holds
for the shifts $T_L^\ast$ in temperature when crossing the phase boundary at fixed
$\Delta$. Higher-order corrections are of the form $V^{-n}=L^{-nD}$ with $n\ge 2$,
where $V$ is the system volume, but exponential corrections can also be relevant for
smaller system sizes \cite{janke1st}. The phase coexistence at the transition point
is connected with the occurrence of a latent heat or latent magnetization that lead
to a divergence of the specific heat $C$ and the magnetic susceptibility $\chi$,
evaluated at the pseudo-critical point, where both show a pronounced peak:
$C_L^{\ast}=C(\Delta_L^{\ast})\sim L^D$ and
$\chi_L^{\ast}=\chi(\Delta_L^{\ast})\sim L^D$.

Above the tricritical temperature $T>T_{\rm t}$, or for crystal fields $\Delta <
\Delta_{\rm t}$,
the model exhibits a {\em second-order} phase transition. This segment of the phase
boundary is expected to be in the Ising universality class~\cite{beale1986}. The
shifts of pseudo-critical points hence follow~\cite{LandBind00}
\begin{equation}\label{eqShiftSecond}
  \Delta_{L}^{\ast} = \Delta_{\rm c} + a L^{-1/\nu},
\end{equation}
where $\nu$ is the critical exponent of the correlation length. An analogous
expression can again be written down for the case of crossing the phase boundary at
constant $\Delta$. The relevant exponents for the Ising universality class are the
well-known Onsager ones, i.e., $\alpha = 0$, $\beta = 1/8$, $\gamma = 7/4$, and
$\nu = 1$. Corrections to the form \eqref{eqShiftSecond} can include analytic and
confluent terms, for a discussion see, e.g., Ref.~\cite{cardy_book}. Since $\alpha=0$
we expect a merely logarithmic divergence of the specific-heat peaks,
$C_L^{\ast}\sim\ln L$. The peaks of the magnetic susceptibility should scale as
$\chi_L^{\ast} \sim L^{\gamma/\nu}$.
%
We recall that critical exponents are not the only universal quantities
\cite{cardy_book}, as these are accompanied by critical amplitude ratios such as the
ratio $U_\xi = f^+/f^-$ of the amplitude of the correlation length scaling
$\xi \sim f^\pm t^{-\nu}$ above and below the critical point
\cite{pelissetto2002}. Less universal are dimensionless quantities in finite-size
scaling such as the ratio of the correlation length and the system size, $\xi/L$,
which for Ising spins on $L\times L$ patches of the square lattice with periodic
boundary conditions for $L\to\infty$ approaches the value \cite{salas2000}
\begin{equation}
  (\xi/L)_\infty = 0.905\,048\,829\,2(4).
  \label{eq:ratio-exact}
\end{equation}
We will study this ratio below for the present system in the second-order regime.
Another weakly universal quantity is the fourth-order magnetization cumulant (Binder
parameter) $V_4$ at criticality \cite{selke2005,pelissetto2002}.

\section{Simulation methods and observables}
\label{secNumerical}

For the present study we used a combination of two advanced simulational setups.  The
bulk of our simulations are performed using a generalized parallel implementation of
the multicanonical approach.  Comparison tests and illustrations of universality are
conducted via a hybrid updating scheme combining Metropolis and generalized Wolff
cluster updates. The multicanonical approach is particularly well suited for the
first-order transition regime of the phase diagram and enables us to sample a broad
parameter range (temperature or crystal field). It also yields decent estimates for
the transition fields in the second-order regime and the corresponding quantities of
interest. For such continuous transitions, the hybrid approach may then be applied
subsequently in the vicinity of the already located pseudo-critical points in order to
obtain results of higher accuracy.  In all our simulations we keep a constant
temperature and cross the phase boundary along the crystal-field axis, in analogy to
our recent study in three dimensions~\cite{zierenberg2015}.

\subsection{Parallel multicanonical approach}
\label{secMuca}

The original multicanonical (muca) method~\cite{berg1991,janke1992} introduces a
correction function to the canonical Boltzmann weight $\exp(-\beta E)$, where
$\beta = 1/(k_{\rm B}T)$ and $E$ is the energy, that is designed to produce a flat
histogram after iterative modification. This can be interpreted as a generalized
ensemble over the phase space $\{\phi\}$ of configurations ($\{\phi\} = \{\sigma_i\}$
for the BC model) with weight function $W[{\cal H}(\{\phi\})]$, where ${\cal H}$ is
the Hamiltonian and $E = {\cal H}(\{\phi\})$. The corresponding generalized partition
function is
\begin{equation}
  Z_{\rm muca} = \int_{\{\phi\}} W[{\cal H}(\{\phi\})]\,\mathrm{d}\{\phi\} = \int \Omega(E)W(E)\,\mathrm{d}E.
\end{equation}
As the second form shows, a flat energy distribution
$P_\mathrm{muca}(E)=\Omega(E)W(E)/Z_\mathrm{muca}={\rm const}.$ is achieved if
$W(E)\propto\Omega^{-1}(E)$, i.e., if the weight is inversely proportional to the
density of states $\Omega(E)$. For the weight function $W^{(n)}(E)$ in iteration $n$,
the resulting normalized energy histogram satisfies
$\langle H^{(n)}(E)\rangle = P^{(n)}(E) = \Omega(E) W^{(n)}(E)/Z^{(n)}$. This
suggests to choose as weight function $W^{(n+1)}(E)=W^{(n)}(E)/H^{(n)}(E)$ for the
next iteration, thus iteratively approaching $W(E) \propto \Omega^{-1}(E)$. In each
of the ensembles defined by $W^{(n)}$, we can still estimate canonical expectation
values of observables $O=O(\{\phi\})$ without systematic deviations as
\begin{equation}
  \langle O\rangle_{\beta}
  =\frac{\langle O(\{\phi\}) e^{-\beta {\cal H}(\{\phi\})}W^{-1}[ {\cal H}(\{\phi\})] \rangle_{\rm muca}}
  {\langle   e^{-\beta {\cal H}(\{\phi\}) }W^{-1}[ {\cal H}(\{\phi\})] \rangle_{\rm muca}}.
  \label{eq:muca-reweight}
\end{equation}

For the present problem we apply the generalized ensemble approach to the
crystal-field component $E_\Delta$ of the energy only, thus allowing us to
continuously reweight to arbitrary values of $\Delta$~\cite{zierenberg2015}. To this
end, we fix the temperature and apply a generalized configurational weight according
to
\begin{equation}
  e^{-\beta(E_{J}+\Delta \Emuca)}
  \rightarrow e^{-\beta E_{J}}~W\left(\Emuca\right).
\end{equation}
The procedure of weight iteration is applied in exactly the same way as before.  Data
from a final production run with fixed $W(\Emuca)$ may be reweighted to the canonical
ensemble via a generalization of Eq.~\eqref{eq:muca-reweight},
\begin{equation}\label{eqReweightDelta}
  \langle O \rangle_{\beta,\Delta}
  =\frac{\langle O e^{-\beta\Delta\Emuca}W^{-1}(\Emuca) \rangle_{\rm muca}}
        {\langle   e^{-\beta\Delta\Emuca}W^{-1}(\Emuca) \rangle_{\rm muca}}.
\end{equation}

As was demonstrated in Ref.~\cite{zierenberg2013}, the multicanonical weight
iteration and production run can be efficiently implemented in a parallel fashion. To
this end, parallel Markov chains sample independently with the same fixed weight
function $W^{(n)}(\Emuca)$. After each iteration, the histograms are summed up and
form independent contributions to the probability distribution
$H^{(n)}(\Emuca)=\sum_i H^{(n)}_i(\Emuca)$. In the present case, we ran our
simulations with $64$ parallel threads and demanded a flat histogram in the range
$\Emuca\in[0,V]$ with a total of $200$ transits, where $V = L^2$ is the total number
of lattice sites. A transit was here defined as a single Markov chain traveling from
one energy boundary to another.

Using this parallelized multicanonical scheme we performed simulations at various
fixed temperatures, cf.\ the data collected below in Table~\ref{tabResults} in the
summary section. For each $T$, we simulated system sizes up to $L_{\rm max} = 128$ in
the second-order regime of the phase diagram ($T>T_{\rm t}$) and up to
$L_{\rm max} \leq 96$ (depending on the temperature) in the first-order regime
($T<T_{\rm t}$). At one particular temperature, namely $T = 1.398$, we were able to
compare with several previous, in part contradictory,
studies~\cite{beale1986,silva2006,malakis}.

\subsection{Hybrid approach}
\label{secHybrid}

For the second-order regime of the phase boundary, our simulations need to cope with
the critical slowing down effect that is not explicitly removed by the multicanonical
approach. Here, we make use of a suitably constructed cluster-update algorithm to
achieve precise estimates close to criticality. While the Fortuin-Kasteleyn
representation of the Ising and Potts models \cite{Fortuin1972} allows for a drastic
reduction or, in some cases, removal of critical slowing down using cluster updates
\cite{swendsen:87a}, the situation is more involved for the BC model, where no
complete transformation to a dual bond language is available. As suggested previously
in Refs.~\cite{Blote95,hasen2010,malakis12}, we therefore rely on a partial
transformation, applying a cluster update only to the spins in the $\pm 1$ states,
ignoring the diluted sites with $\sigma_i = 0$. This approach alone is clearly not
ergodic as the number and location of $\sigma_i = 0$ sites is invariant, and we hence
supplement it by a local Metropolis update. For the cluster update of the $\pm 1$
spins we use the single-cluster algorithm due to Wolff \cite{Wolff1989}.  In the
present hybrid approach an elementary MC step (MCS) is the following heuristically
determined mixture: after each Wolff step we attempt $3\times L$ Metropolis spin
flips and the elementary step consists of $L$ such combinations. In other words, a
MCS has $3$ Metropolis sweeps and $L$ Wolff steps.

The convergence of the hybrid approach may be easily checked for every lattice size
used in the simulations. For instance, to observe convergence for $L=24$ we used $3$
different runs consisting of $12800\times V$, $25600\times V$, and $51200\times V$
(about $30\times 10^6$) MCS, whereas for $L=48$ we compared another set of $3$
different runs consisting of $2560\times V$, $5120\times V$, and $10240\times V$
(about $23\times 10^6$) MCS. In all our simulations a first large number of MCS was
disregarded until the system was well equilibrated. Our runs using this technique
covered a range of different temperatures in the second-order regime and, in
particular, the selected temperature $T = 1.398$ mentioned above, using system sizes
up to $L_{\rm max} = 128$. For each $L$, we used up to $100$ independent runs
performed in parallel to increase our statistical accuracy.

\begin{figure}
  \hspace{0.2cm}(a)\hspace{5.8cm}  (b)
  \vspace{-1.5em}
  \begin{center}
    \includegraphics{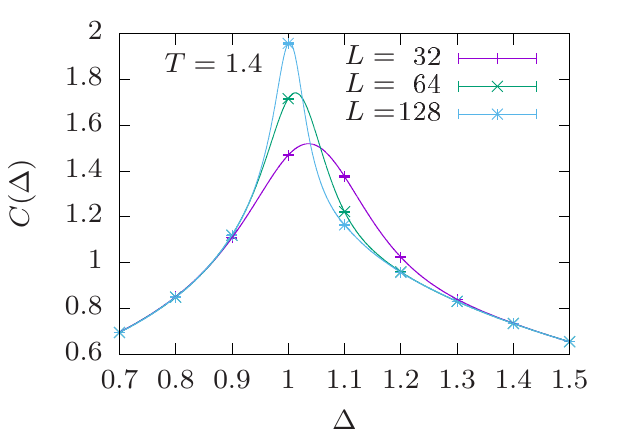}
    \includegraphics{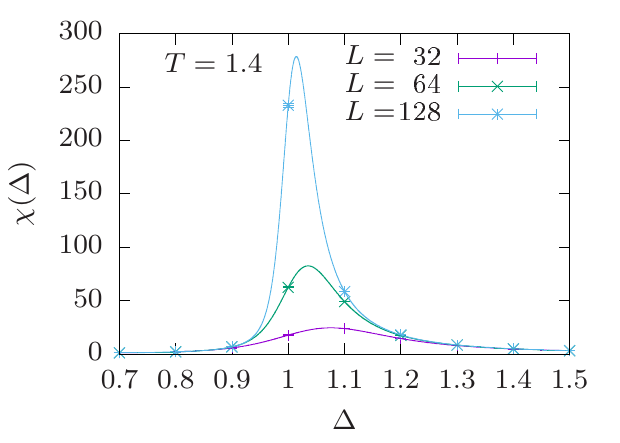}
  \end{center}
  \caption{\label{figCChi} Specific heat (a) and susceptibility (b) curves as a
    function of the crystal field $\Delta$ obtained from parallel multicanonical
    simulations at $T = 1.4$ The lines show simulation results as continuous
    functions of $\Delta$ from reweighting, the individual points indicate the size
    of statistical errors.}
\end{figure}

\subsection{Observables}
\label{secObservables}

For the purpose of the present study we focused on the specific heat, the magnetic
susceptibility and the correlation length. As in our simulations we cross the
transition line at fixed temperature, it is reasonable to study the crystal-field
derivative $\partial \langle E\rangle/\partial \Delta$ instead of the temperature
gradient $\partial \langle E\rangle/\partial T$. As was pointed out in
Ref.~\cite{zierenberg2015}, the singular behavior is also captured in the simpler
quantity
\begin{equation}\label{eqC}
  C \equiv \frac{\partial \langle E_J\rangle}{\partial\Delta}\frac{1}{V}
  =  - \beta \left(\langle E_J\Emuca\rangle - \langle
  E_J\rangle\langle\Emuca\rangle\right)/V.
\end{equation}
The magnetic susceptibility is defined as the field derivative of the
absolute magnetization, and this yields
\begin{equation}\label{eqChi}
  \chi = \beta\left(\langle M^2\rangle-\langle |M|\rangle^2\right)/V,
\end{equation}
where $M=\sum_i\sigma_i$. As we will discuss below, however, the use of the modulus
$|M|$ to break the symmetry on a finite lattice leads to some subtleties for the BC
model, especially in the first-order regime. Exemplary plots of $C$ and $\chi$ as a
function of the crystal field $\Delta$ obtained from the multicanonical simulations
are shown in Fig.~\ref{figCChi}.  It is obvious from these plots that both
$C(\Delta)$ and $\chi(\Delta)$ show a size-dependent maximum, together with a shift
behavior of peak locations.

Let us define $\Delta^{\ast}_{L,C}$ and $\Delta^{\ast}_{L,\chi}$ as the crystal-field
values which maximize $C(\Delta)$ and $\chi(\Delta)$, respectively. These are
pseudo-critical points that should scale according to Eqs.~\eqref{eqShiftFirst} and
\eqref{eqShiftSecond}, respectively.
They are numerically determined by a bisection algorithm that iteratively performs
histogram reweighting in the vicinity of the peak, detecting the point of locally
vanishing slope. Error bars are obtained by repeating this procedure for 32 jackknife
blocks~\cite{efron1982}.  Similarly, we denote by $C^{\ast}_L=C(\Delta^{\ast}_{L,C})$
and $\chi^{\ast}_L=\chi(\Delta^{\ast}_{L,\chi})$ the values of the specific heat and
the magnetic susceptibility at their pseudo-critical points, respectively. These may
be directly evaluated as canonical expectation values according to
Eq.~(\ref{eqReweightDelta}).

We finally also studied the second-moment
correlation length $\xi$~\cite{cooper1989,ballesteros2001}. This involves the
Fourier transform of the spin field $\hat{\sigma}(\mathbf{k}) = \sum_{\bf
x}\sigma_{\bf x}e^{i{\bf kx}}$. If we set
{$F = \left\langle |\hat{\sigma}(2\pi/L,0)|^2+|\hat{\sigma}(0,2\pi/L)|^2\right\rangle/2$},
the correlation length can be obtained via~\cite{ballesteros2001}
\begin{equation}
 \xi  \equiv  \frac{1}{2\sin(\pi/L)}\sqrt{\frac{\langle M^2\rangle}{F}-1}.
 \label{eq:xi}
\end{equation}
From $\xi$ we may compute the ratio $\xi/L$, which tends to a weakly universal
constant for $L\to\infty$ as discussed above in Sec.~\ref{secLiterature}.

\section{Numerical results}
\label{secFSS}

\begin{figure}
  \centering
  \includegraphics{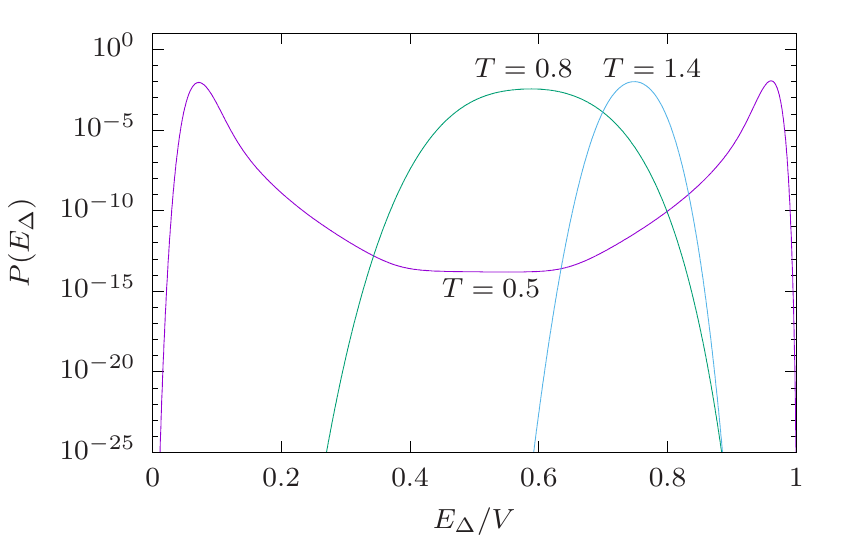}
  \caption{\label{figProbDist}%
    Canonical probability distribution $P(E_{\Delta})$ at the transition field
    $\Delta_{L,C}^\ast$ for various temperatures for $L=48$. Note the logarithmic scale
    on the vertical axis.  }
\end{figure}

In this section we present our main finite-size scaling analysis, covering both
first- and second-order transition regimes of the phase diagram of the square-lattice
model. We begin by presenting the canonical probability distribution $P(\Emuca)$ at
the pseudo-critical crystal fields $\Delta^{\ast}_{L,C}$ for different temperatures.
Figure~\ref{figProbDist} shows $P(\Emuca)$ for the temperature $T=0.5$, which is in
the first-order regime, and for $T=0.8$ and $1.4$, which are in the second-order
regime of the transition line, for a system size $L=48$.  Well inside the first-order
transition regime the system shows a strong suppression of transition states,
connected to a barrier between two coexisting phases. This is characteristic of a
discontinuous transition. Here, the barrier separates a spin-$0$ dominated (small
$\Emuca$, $\Delta>\Delta^{\ast}$) and a spin-$\pm 1$ dominated (large $\Emuca$,
$\Delta<\Delta^{\ast}$) phase. In this regime, the model qualitatively describes the
superfluid transition in $^3$He-$^4$He mixtures.  As the temperature increases and
exceeds the tricritical point $T_{\rm t}\approx 0.608$, the barrier disappears and
the probability distribution shows a unimodal shape, characteristic of a continuous
transition. In this regime the model qualitatively describes the lambda line of
$^3$He-$^4$He mixtures.

\paragraph{First-order regime:}

\begin{figure}[t]
  \hspace{0.2cm}(a)\hspace{5.8cm}  (b)
  \vspace{-1.5em}
  \begin{center}
    \includegraphics{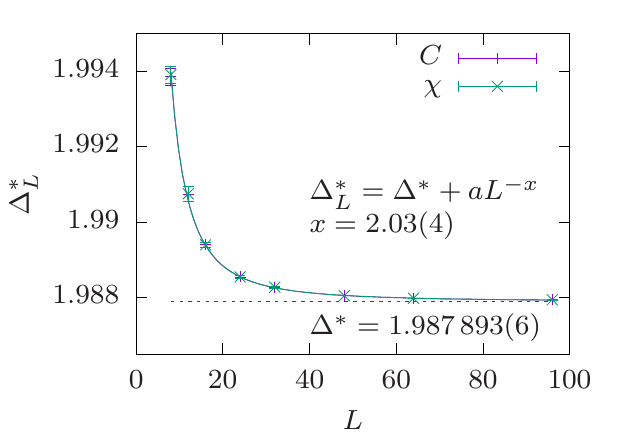}
    \includegraphics{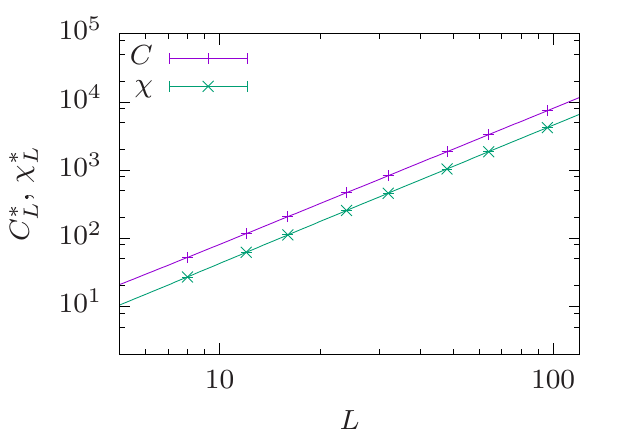}
  \end{center}
  \caption{\label{figScalingT0.5}%
    Finite-size scaling analysis in the first-order transition regime ($T=0.5$) based
    on the specific heat and magnetic susceptibility. (a) Simultaneous fit of the
    functional form \eqref{eq:field-fit-form} to the pseudo-critical fields of the
    specific heat and susceptibility. (b) Scaling of the values $C_L^\ast$ and
    $\chi_L^\ast$ at these maxima together with fits of the form \eqref{eq:cscaling}
    and \eqref{eq:chiscaling} to the data.  }
\end{figure}

Here we focus on one particular temperature, namely $T=0.5$, to verify the expected
scaling discussed in Sec.~\ref{secLiterature}.  Figure~\ref{figScalingT0.5}(a) shows
a finite-size scaling analysis of the pseudo-critical fields for which we expect
shifts of the form
\begin{equation}
   \Delta^{\ast}_{L,O} = \Delta^{\ast} + a_O L^{-x}.
   \label{eq:field-fit-form}
\end{equation}
We performed simultaneous fits to $\Delta^{\ast}_{L,C}$ and $\Delta^{\ast}_{L,\chi}$
with a common value of $x$. Including the full range of data $L=8-96$ we obtain
$\Delta^{\ast}=1.987\,893(6)$ and $x=2.03(4)$ with $Q\approx0.98$.\footnote{$Q$ is
  the probability that a $\chi^2$ as poor as the one observed could have occurred by
  chance, i.e., through random fluctuations, although the model is correct
  \cite{young2015}.} This is consistent with the most recent and very precise
estimate $\Delta^{\ast} = 1.987\,89(1)$ by Kwak et al.~\cite{kwak2015}, and the
theoretical prediction $x = D = 2$. We note that for all fits performed here, we
chose a minimum system size to include in the fit such that a goodness-of-fit
parameter $Q>0.1$ was achieved. For the specific heat at the maxima, we expect the
leading behavior $C^{\ast}_L \sim L^D$. Scaling corrections at first-order
transitions are in inverse integer powers of the volume \cite{janke1st}, $L^{-nD}$,
$n=1$, $2$, $\ldots$, so we attempted the fit form
\begin{equation}
  C^{\ast}_L         = b_C L^{x}   \left(1+b^\prime_C L^{-2}\right)
  \label{eq:cscaling}
\end{equation}
and we indeed find an excellent fit for the full range of system sizes, yielding
$x=1.9999(2)$ and $Q=0.78$ --- note that due to the value of $x$, the $1/L^2$
correction simply corresponds to an additive constant. The amplitudes are
$b_C = 0.8065(6)$ and $b^\prime_C = 0.84(5)$. This fit and the corresponding data are
shown in Fig.~\ref{figScalingT0.5}(b).

\begin{figure}
  \hspace{0.2cm}(a)\hspace{5.8cm}  (b)
  \vspace{-1.5em}
  \begin{center}
    \includegraphics{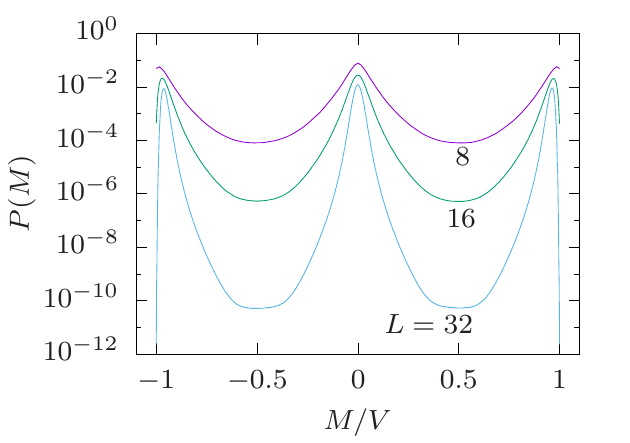}
    \includegraphics{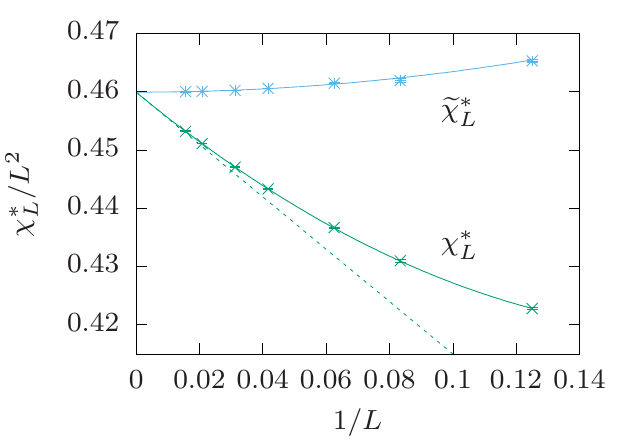}
  \end{center}
  \caption{\label{figScalingT0.5chiM2}%
    (a) Canonical probability distribution $P(M)$ of the magnetization at $T=0.5$ and
    the pseudo-critical point $\Delta_{L,\chi}^{\ast}$ for selected lattice sizes. The
    middle peak corresponds to the disordered phase with a majority of $\sigma_i = 0$
    states, while the peaks on the left and right stem from the ordered phases with
    pre-dominance of $\sigma_i = \pm 1$. (b) Corrections to the finite-size scaling
    of the magnetic susceptibility $\chi$ of Eq.~\eqref{eqChi} and to
    $\widetilde{\chi} = \beta(\langle\widetilde{M}^2\rangle -
    \langle\widetilde{M}\rangle^2)/V$
    with $\widetilde{M}$ according to Eq.~(\ref{eq:mtilde}). The solid lines show a
    fit of the form $a+b/L+c/L^2$, including an inversely linear term, for $\chi$
    ($Q=0.85$) and a fit of the form $a+b/L^2$ for $\widetilde{\chi}$ ($Q=0.30$).
    Both quantities are evaluated at the locations $\Delta_{L,\chi}^{\ast}$ of the
    maxima of $\chi$.  }
\end{figure}

For the magnetic susceptibility, on the other hand, the correction proportional
to $L^{-2}$ is not sufficient to describe the data down to small $L$, and
neither are higher orders $L^{-4}$, $L^{-6}$ etc. We also experimentally
included an exponential correction which is expected to occur in the first-order
scenario and occasionally can be relevant for small $L$ \cite{janke1st}, but
this also did not lead to particularly good fits. Using an additional $1/L$
correction, on the other hand, i.e., a fit form
\begin{equation}
  \chi^{\ast}_L      = b_\chi L^{x}\left(1+b^\prime_\chi L^{-1}+b^{\prime\prime}_\chi  L^{-2}\right)
  \label{eq:chiscaling}
\end{equation}
yields excellent results with $Q=0.98$ and $x=2.001(1)$ for the full range
$L=8$--$96$ of system sizes, the corresponding fit and data are also shown in
Fig.~\ref{figScalingT0.5}(b). Here, $b_\chi = 0.458(2)$, $b^\prime_\chi = -0.94(5)$
and $b^{\prime\prime}_\chi = 2.5(3)$. A $1/L$ correction term is not expected at a
first-order transition \cite{janke1st}, but its presence is rather clear from our
data. Some further consideration reveals that it is, in fact, an artifact resulting
from the use of the modulus $|M|$ in defining $\chi$ in Eq.~\eqref{eqChi}. To see
this, consider the shape of the magnetization distribution function at the transition
point $\Delta^\ast_{L,\chi}$ shown for different system sizes in
Fig.~\ref{figScalingT0.5chiM2}(a). The middle peak corresponds to the disordered
phase dominated by $0$-spins, while the peaks on the left and right represent the
ordered $\pm 1$ phases. While in $P(M)$, the middle peak is symmetric around zero and
hence $\langle M\rangle_d = 0$ in the disordered phase, the modulus $|M|$ will lead
to an average $\langle |M|\rangle_d = O(L)$ of the order of the peak
width\footnote{The width of the peak is estimated from the fact that $O(V)$ spins in
  the disordered phase equal $+1$ and $O(V)$ others equal $-1$. Hence their sum is of
  order $O(\sqrt{V}) = O(L)$.}. Since $\chi$ measures the square width of the
distribution of $|M|$, this will have a $1/L$ correction stemming from this $O(L)$
contribution to $|M|$.

This problem can be avoided by employing a different method of breaking the symmetry
on a finite lattice. One possible definition could be
\begin{equation}
  \label{eq:mtilde}
  \widetilde{M} = \left\{
    \begin{array}{l@{\hspace{0.75em}}l}
      M & \mbox{for}\;\;|M|/V < 0.5 \\
      |M| & \mbox{for}\;\;|M|/V \ge 0.5
    \end{array}
  \right.,
\end{equation}
which only folds the $-1$-peak onto the $+1$-peak, but leaves the $0$-peak untouched.
As is seen from the data and fits shown in Fig.~\ref{figScalingT0.5chiM2}(b) in
contrast to $\chi$ the scaling of the corresponding susceptibility
$\widetilde{\chi} = \beta(\langle\widetilde{M}^2\rangle -
\langle\widetilde{M}\rangle^2)/V$
does not show a $1/L$ correction, but only the volume correction $\propto 1/L^2$
expected for a first-order transition.

Overall, it is apparent that our simulations nicely reproduce the behavior expected
for a first-order transition, whereas a conventional canonical-ensemble simulation
scheme would be hampered by metastability and hyper-critical slowing down.

\paragraph{Second-order regime:}

\begin{figure}[t]
  \hspace{0.2cm}(a)\hspace{5.8cm}  (b)
  \vspace{-1.5em}
  \begin{center}
    \includegraphics{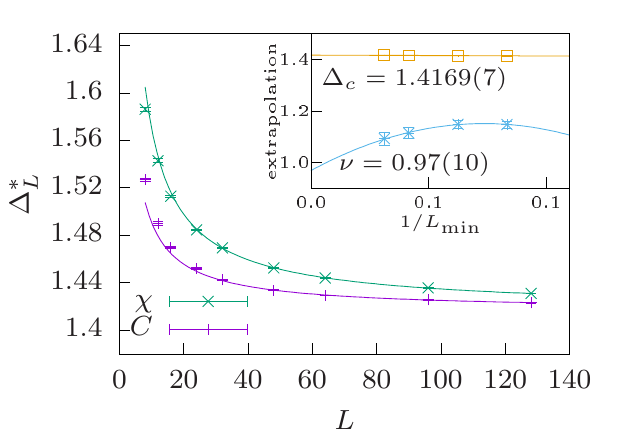}
    \includegraphics{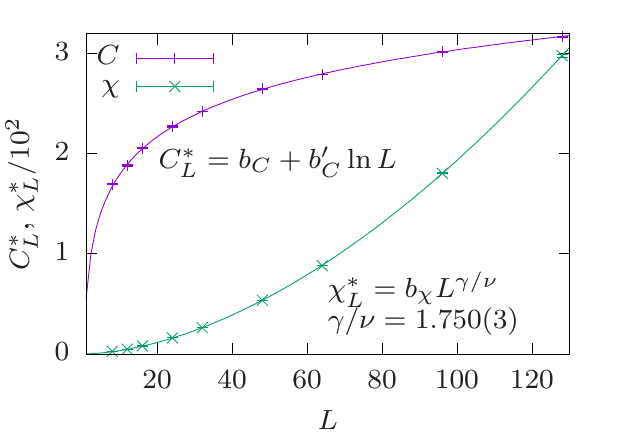}
  \end{center}
  \caption{\label{figScalingT1.2}%
    Finite-size scaling analysis in the second-order transition regime ($T=1.2$)
    based on the specific heat and magnetic susceptibility from data of the
    multicanonical simulations. Panel (a) shows a
    simultaneous fit of the pseudo-critical fields to the leading-order ansatz
    (\ref{eq:second-order-fits1}). The effective estimates are further subjected to
    an extrapolation in $1/L_\mathrm{min}$ as shown in the inset ($\Delta_{\rm eff}$
    linear and $\nu_{\rm eff}$ quadratic in the lower-fit bound $1/L_{\rm
      min}$).
    Panel (b) shows fits of the maxima at the pseudo-critical points with the
    predicted behavior. The results verify the expected Ising universality class.  }
\end{figure}

We continue with the second-order regime, again focusing on one particular
temperature, $T=1.2$, in order to verify the expected scaling as discussed in
Sec.~\ref{secLiterature}. We restrict ourselves here to the leading-order scaling
expressions,
\begin{eqnarray}
  \Delta^{\ast}_{L,O} &=& \Delta_{\rm eff} + a_O L^{-1/\nu_{\rm eff}}, \label{eq:second-order-fits1} \\
  C^{\ast}_L          &=& b_C + b_C^\prime \ln L, \label{eq:second-order-fits2} \\
  \chi^{\ast}_L       &=& b_\chi L^{\gamma/\nu}, \label{eq:second-order-fits3}
\end{eqnarray}
taking scaling corrections into account by systematically omitting data from the
small-$L$ side of the full range $L=8-128$.  Figure~\ref{figScalingT1.2} shows the
FSS analysis for $T=1.2$. A simultaneous fit of the pseudo-critical fields in
Fig.~\ref{figScalingT1.2}(a) yields $Q\approx0.26$ for $L\ge 32$ with
$\Delta_{\rm eff}=1.4161(6)$ and
$\nu_{\rm eff}=1.09(3)$. This is only marginally consistent with the expected Ising
value $\nu=1$. We attribute this effect to the presence of scaling corrections. We
hence performed a further finite-size scaling analysis of $\Delta_{\rm eff}$ and
$\nu_{\rm eff}$ as a function of the inverse lower fit range $1/L_{\rm min}$ to
effectively take these corrections into account. For a quadratic fit in $1/L$ for
$\nu_{\rm eff}$ we find $\nu=0.97(10)$, while a linear fit for $\Delta_{\rm eff}$
yields $\Delta_{\rm c}=1.4169(7)$, see the inset of Fig.~\ref{figScalingT1.2}(b).

We further checked for consistency with the Ising universality by considering the
scaling of the maxima of the specific heat and magnetic susceptibility as shown in
Fig.~\ref{figScalingT1.2}(b). The specific heat shows a clear logarithmic scaling
behavior for $L\ge 12$ with $Q\approx0.16$ in strong support of $\alpha=0$. Moreover,
a power-law fit to the magnetic susceptibility peaks yields for $L\geq32$ a value
$\gamma/\nu=1.750(3)$ with $Q\approx0.24$, in perfect agreement with the Ising value
$\gamma/\nu=7/4$. Overall, this reconfirms the Ising universality class, with similar
results for other $T>T_{\rm t}$.

\paragraph{Correlation length:}

\begin{figure}
  \hspace{0.2cm}(a)\hspace{5.8cm}  (b)
  \vspace{-1.5em}
  \begin{center}
    \includegraphics{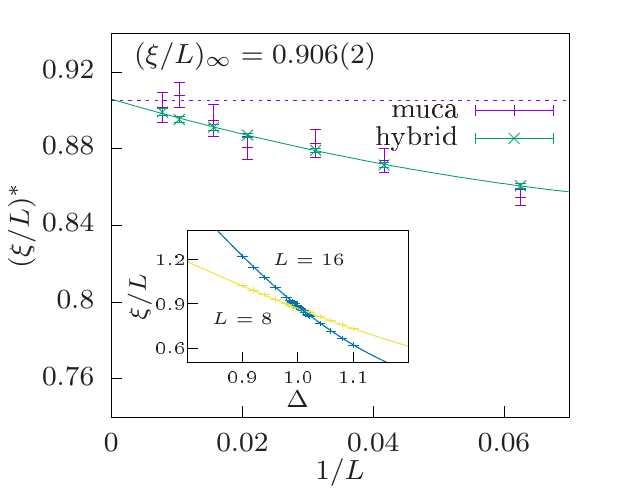}
    \includegraphics{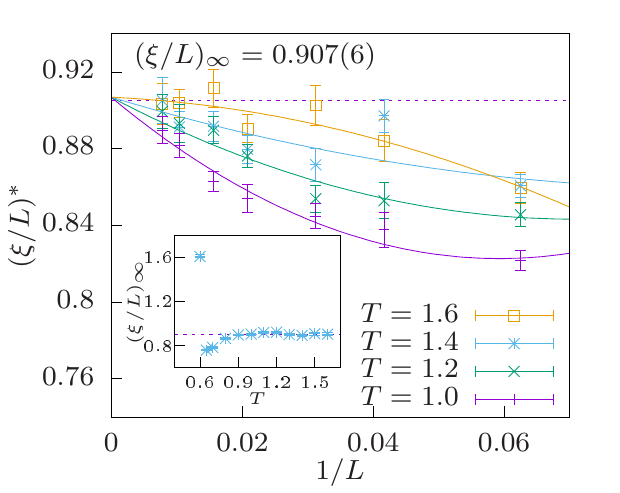}
  \end{center}
  \caption{\label{figXiUniversality}%
    Finite-size scaling of the correlation length crossings $(\xi/L)^{\ast}$. The
    dashed horizontal line in both panels shows the asymptotic value for the
    square-lattice Ising model with periodic boundaries according to
    Eq.~\eqref{eq:ratio-exact}. (a) Results for $T=1.398$, comparing data from the
    multicanonical and hybrid methods. The line shows a quadratic fit in $1/L$ to the
    data from the hybrid method. The inset demonstrates the crossing point of $L=8$
    and $L=16$ from both muca (lines) and hybrid (data points). (b) Simultaneous fit
    for several temperatures obtained from multicanonical simulations.
    The inset shows results from direct fits for a range of temperatures
    in comparison to the asymptotic value of Eq.~\eqref{eq:ratio-exact}. Well above
    the tricritical point, $(\xi/L)^{\ast}$ nicely converges towards the Ising value.
    Towards the tricritical point, additional corrections emerge.  }
\end{figure}

We now turn to a discussion of the correlation length $\xi$. This is where we used
the results of the hybrid method for improved precision. We determined the
second-moment correlation length according to Eq.~\eqref{eq:xi} and then used the
quotient method to determine the limiting value of the ratio
$\xi/L$~\cite{amit05,night,bal96,fytas_PRL}. We define a series of pseudo-critical
points $\Delta_{(L,2L)}^{\ast}$ as the value of the crystal field where
$\xi_{2L}/\xi_L=2$. These are the points where the curves of $\xi/L$ for the sizes
$L$ and $2L$ cross.  A typical illustration of this crossing is shown in the inset of
Fig.~\ref{figXiUniversality}(a) for $T=1.398$.  The pair of system sizes considered
is ($8$, $16$) and the results shown are obtained via both the hybrid method (data
points) and the multicanonical approach through quasi continuous reweighting (lines).
Denote the value of $\xi/L$ at these crossing points as $(\xi/L)^{\ast}$. The size
evolution of $(\xi/L)^{\ast}$ and its extrapolation to the thermodynamic limit,
denoted by $(\xi/L)_{\infty}$, will provide us with the desired test of
universality. In Fig.~\ref{figXiUniversality}(a) we illustrate the
$L\rightarrow \infty$ extra\-polation of $(\xi/L)^{\ast}$ for the previously studied
case $T = 1.398$ and compare the two simulation schemes, hybrid and
multicanonical. The sequence of pairs of system sizes considered is as follows: ($8$,
$16$), ($12$, $24$), ($16$, $32$), ($24$, $48$), ($32$, $64$), ($48$, $96$), and
($64$,$128$). It is seen that the results obtained with the hybrid method suffer less
from statistical fluctuations. It is found that a second-order polynomial in $1/L$
describes the data for $(\xi/L)$ well and a corresponding fit yields
\begin{equation}
  (\xi/L)_{\infty}^{\rm (hybrid)} = 0.906(2).
\end{equation}
A similar fitting attempt to the multicanonical data gives an estimate of
\begin{equation}
  (\xi/L)_{\infty}^{\rm (muca)} = 0.913(9),
\end{equation}
consistent with but less accurate than $(\xi/L)_{\infty}^{\rm (hybrid)}$. Both
estimates are fully consistent with the exact value
$(\xi / L)_{\infty} = 0.905\,048\,8292(4)$~\cite{salas2000}.

In Fig.~\ref{figXiUniversality}(b) we present a complementary illustration using data
from the multicanonical approach and several temperatures in the second-order
transition regime of the phase diagram, as indicated by the different colors. In
particular, we show the values $(\xi/L)^{\ast}$ for several pairs of system sizes
from ($8$, $16$) up to ($64$, $128$). The solid lines are second-order polynomial
fits in $1/L$, imposing a common $L\rightarrow \infty$ extrapolation
$(\xi/L)_{\infty}$.  The result obtained in this way is $0.907(6)$, again well
compatible with the exact Ising value. We note that following the discussion in
Ref.~\cite{salas2000}, a correction with exponent $1/L^2$ or possibly $1/L^{7/4}$ is
expected, but a term proportional to $1/L$ is not. Here, however, we do not find
consistent fits with $1/L^{7/4}$ or $1/L^2$ only, and using a second-order polynomial
in $1/L^w$ instead we find $w=0.91(27)$, consistent with the two terms $1/L$ and
$1/L^2$. A possible explanation for this behavior might be a non-linear dependence of
the scaling fields on $L$ as a linear correction in reduced temperature $t$ produces
a term $L^{-1/\nu}$ in FSS, and $\nu = 1$ \cite{pelissetto2002}. In the inset of
Fig.~\ref{figXiUniversality}(b) we show the values of $(\xi/L)_{\infty}$ for various
further temperatures.  In this case, each estimate of $(\xi/L)_{\infty}$ is obtained
from individual quadratic fits on each data set without imposing a common
thermodynamic limit.  The departure from the Ising value $0.905$, which is again
marked by the dashed line, is clear as $T \rightarrow T_{\rm t}$. There, additional
higher-order corrections due to the crossover to tricritical scaling become relevant.

Finally, we also considered the behavior of the susceptibility and specific heat from
runs of the hybrid method, evaluated at the pseudo-critical points
$\Delta_{(L,2L)}^{\ast}$ from the crossings of $\xi/L$. For $\chi$ we find an
excellent fit for the full range of lattice sizes with the pure power-law form
\eqref{eq:second-order-fits3}, resulting in $\gamma/\nu = 1.75(2)$
($Q=0.96$). Similarly, a fully consistent fit is found over the full lattice size
range for the specific heat using the logarithmic form \eqref{eq:second-order-fits2}
($Q=1.0$).

\paragraph{Full temperature range:}

Having established the common first-order scaling for $T<T_{\rm t}$ and the Ising
universality class for $T>T_{\rm t}$, we attempted to improve the precision in the location
of the phase boundary for the square lattice model. To this end, we considered
simultaneous fits of the scaling ans\"atze Eqs.~(\ref{eqShiftFirst}) and
(\ref{eqShiftSecond}) to the peak locations $\Delta^{\ast}_{L,C}$ and
$\Delta^{\ast}_{L,\chi}$, depending on whether the considered temperature is in the
first-order or in the second-order regime,
\begin{eqnarray}
  \Delta^{\ast}_{L,O} &=& \Delta^{\ast} + a_O L^{-D}      \hspace{5.0em}\text{for
                          $T<T_{\rm t}$}, \\
  \Delta^{\ast}_{L,O} &=& \Delta_{\rm c} + a_O L^{-1/\nu} \hspace{4.5em}\text{for
  $T>T_{\rm t}$},
\end{eqnarray}
with $D=2$ and $\nu=1$ fixed. As before, we take corrections to scaling into account
by systematically omitting data from the small-$L$ side until fit qualities $Q>0.1$
are achieved. The results for the transition fields are listed in
Table~\ref{tabResults}, including fit errors.  Well inside the first-order regime,
fits are excellent and cover the full data set ($L\geq 8$), so scaling corrections
are not important there. Around the tricritical point, fits become difficult. For
example, a simultaneous fit for $T=0.65$ with $L\geq64$ still yields
$Q\approx0.07$. This is, of course, no surprise as we should see a crossover to the
tricritical scaling there. Moving away from the tricritical point into the
second-order regime, fits become more feasible. Corrections appear to be smallest
between $T=0.9$ and $T=1.0$ where we could include the full data set, $L\geq8$, with
$Q\approx0.4$ each. Increasing the temperature, we then again find stronger
corrections. Particularly for $T=1.6$ fits with $L\geq64$ are required to obtain
$Q>0.1$. We attribute this effect to the fact that our variation of $\Delta$ is
almost tangential to the phase boundary there, so field-mixing effects should be
quite strong
\cite{wilding1996}. Overall, we find very good agreement with recent previous
studies, but often increased precision, cf.\ the data in Table~\ref{tabResults}.

\begin{table}[tb]
\caption{%
  Representative points in the phase diagram of the Blume-Capel model on the
  square lattice from previous studies and the present work. In the first two
  columns we either indicate the value of $\Delta$ for simulations that vary $T$
  or the value of $T$ for simulations that vary $\Delta$. Error bars are given
  in parenthesis in either $\Delta$ or $T$, depending on the simulation type.
}
\label{tabResults}
\begin{tabular}{ll|llllll}
\hline\hline\noalign{\smallskip}
& & \multicolumn{2}{l}{Beale}             & Silva et al.      & Malakis et al.      & Kwak et al.      & This work\\
& & \multicolumn{2}{l}{Ref.~\cite{beale1986}} & Ref.~\cite{silva2006} & Ref.~\cite{malakis} & Ref.~\cite{kwak2015} &          \\
\noalign{\smallskip}\hline\noalign{\smallskip}

$\Delta$ & $T$ & $\Delta$ & $T$ & $T$ & $T$ & $\Delta$ & $\Delta$ \\

\noalign{\smallskip}\hline\noalign{\smallskip}
 0    &      &          & 1.695& 1.714(2)& 1.693(3)&               &            \\
      & 1.6  &          &      &         &         &               & 0.375(2)   \\
      & 1.5  &          &      &         &         &               & 0.7101(5)  \\
 0.5  &      &          & 1.567& 1.584(1)& 1.564(3)&               &            \\
      & 1.4  &          &      &         &         &               & 0.9909(4)  \\
      & 1.398&          &      &         &         &               & 0.9958(4)  \\
 1.0  &      &          & 1.398& 1.413(1)& 1.398(2)&               &            \\
      & 1.3  &          &      &         &         &               & 1.2242(4)  \\
      & 1.2  &          &      &         &         &               & 1.4167(2)  \\
 1.5  &      &          & 1.15 & 1.155(1)& 1.151(1)&               &            \\
      & 1.1  &          &      &         &         &               & 1.5750(2)  \\
      & 1.0  &          &      &         &         &               & 1.70258(7) \\
 1.75 &      &          &      &         & 0.958(1)&               &            \\
      & 0.9  &          &      &         &         &               & 1.80280(6) \\
      & 0.8  & 1.87     &      &         &         &               & 1.87879(3) \\
 1.9  &      &          &      & 0.755(3)& 0.769(1)&               &            \\
      & 0.7  & 1.92     &      &         &         &               & 1.93296(2) \\
 1.95 &      &          &      & 0.651(2)& 0.659(2)&               &            \\
      & 0.65 & 1.95     &      &         &         & 1.9534 (1)    & 1.95273(1) \\
      & 0.61 & 1.9655   &      &         &         &               &            \\
      & 0.608&          &      &         &         & 1.96604 (1)   &            \\
      & 0.6  & 1.969    &      &         &         & 1.96825 (1)   & 1.968174(3)\\
 1.975&      &          &      &         & 0.574(2)&               &            \\
 1.992&      &          &      & 0.499(3)&         &               &            \\
      & 0.5  & 1.992    &      &         &         & 1.98789 (1)   & 1.987889(5)\\
      & 0.4  &          &      &         &         & 1.99681 (1)   & 1.99683(2) \\
\noalign{\smallskip}\hline\hline
\end{tabular}
\end{table}

\section{Summary and outlook}
\label{secSummary}

In this paper we have reviewed and extended the phase diagram of the 2D Blume-Capel
model in the absence of an external field, providing extensive numerical results for
the model on the square lattice. In particular, we studied in some detail the
universal ratio $\xi/L$ that allows to confirm the Ising universality class of the
model in the second-order regime of the phase boundary.
In contrast to most previous work, we focused on crossing the phase boundary at
constant temperature by varying the crystal field
$\Delta$~\cite{zierenberg2015}. Employing a multicanonical scheme in $\Delta$ allowed
us to get results as continuous functions of $\Delta$ and to overcome the free-energy
barrier in the first-order regime of transitions. A finite-size scaling analysis
based on a specific-heat-like quantity and the magnetic susceptibility provided us
with precise estimates for the transition points in both regimes of the phase diagram
that compare very well to the most accurate estimates of the current literature.
We have been able to probe the first-order nature of the transition in the
low-temperature phase and to illustrate the Ising universality class in the
second-order regime of the phase diagram. We are also able to provide accurate
estimates of the critical exponents $\nu$ and $\gamma/\nu$, as well as to clearly
confirm the logarithmic divergence of the specific-heat peaks. Using additional
simulations based on a hybrid cluster-update approach we studied the correlation
length in the second-order regime. Via a detailed scaling analysis of the universal
ratio $\xi/L$, we could show that it cleanly approaches the value
$(\xi/L)_\infty = 0.905\ldots$ of the Ising universality class for all temperatures
up to the tricritical point.

In the first-order regime we found a somewhat surprising $1/L$ correction in the
scaling of the conventional susceptibility defined according to Eq.~\eqref{eqChi}. As
it turns out, this is due to the explicit symmetry breaking by using $|M|$ instead of
$M$ in the definition of $\chi$. For a modified symmetry breaking prescription that
leaves the disordered peak invariant, this correction disappears. It would be
interesting to see whether similar corrections are found in other systems with
first-order transitions, such as the Potts model.

To conclude, the Blume-Capel model serves as an extremely useful prototype system for
the study of phase transitions, exhibiting lines of second-order and first-order
transitions that meet in a tricritical point. Apart from the interest in tricritical
scaling, this model hence also allows to investigate the effect of disorder on phase
transitions of different order within the same model. A study of the disordered
version of the model is thus hoped to shed some light on questions of universality
between the continuous transitions in the disordered case that correspond to
different transition orders in the pure model~\cite{fytas_PRE}.

\acknowledgement{ The article is dedicated to Wolfhard Janke on the occasion of his
  60th birthday. M.W.\ thanks Francesco Parisen Toldin for enlightening discussions
  on corrections to finite-size scaling. N.G.F.\ and M.W.\ are grateful to Coventry
  University for providing Research Sabbatical Fellowships that supported part of
  this work. The project was in part funded by the Deutsche Forschungsgemeinschaft
  (DFG) under Grant No.\ JA~483/31-1, and financially supported by the
  Deutsch-Franz\"osische Hochschule (DFH-UFA) through the Doctoral College
  ``${\mathbb L}^4$'' under Grant No.\ CDFA-02-07 as well as by the EU FP7 IRSES
  network DIONICOS under contract No.\ PIRSES-GA-2013-612707.  N.G.F.\ would like to
  thank the Leipzig group for its hospitality during several visits over the last
  years where part of this work was initiated.  }


\end{document}